\documentclass[10pt,twocolumn,twoside]{IEEEtran}

\ifCLASSINFOpdf
  \usepackage[pdftex]{graphicx}
\else
\fi

\usepackage{times}
\usepackage{epsfig}
\usepackage{amsmath}
\usepackage{amssymb}
\usepackage{wasysym}
\usepackage{color}
\usepackage{multirow}
\usepackage{cite}
\usepackage{graphicx} 
\usepackage{subfigure}

\usepackage{algorithm}
\usepackage{algorithmic}
\newcommand{\ie}{{\it i.e.}}
\newcommand{\eg}{{\it e.g.}}

\begin{document}
%
\title{Query-Adaptive Hash Code Ranking for Large-Scale Multi-View Visual Search}

\author{
Xianglong~Liu,~\IEEEmembership{Member,~IEEE,}
Lei~Huang,~
Cheng~Deng,~\IEEEmembership{Member,~IEEE,}
Bo~Lang,~
Dacheng~Tao,~\IEEEmembership{Fellow,~IEEE}
\thanks{X. Liu, L. Huang and B. Lang are with the State Key Lab of Software Development Environment, Beihang University, Beijing 100191, China (email: xlliu@nlsde.buaa.edu.cn).}
\thanks{C. Deng is with School of Electronic Engineering, Xidian University, Xi an 710071, China (email: chdeng.xd@gmail.com)}
\thanks{D. Tao (corresponding author) is with the Centre for Quantum Computation and Intelligent Systems, Faculty of Engineering and Information Technology, University of Technology at Sydney, Sydney, NSW 2007, Australia (e-mail: dacheng.tao@uts.edu.au).}
\thanks{$\copyright$ 20XX IEEE. Personal use of this material is permitted. Permission from IEEE must be obtained for all other uses, in any current or future media, including reprinting/republishing this material for advertising or promotional purposes, creating new collective works, for resale or redistribution to servers or lists, or reuse of any copyrighted component of this work in other works.}

}

\markboth{IEEE Transactions on XXX,~Vol.~X, No.~X, XXX~2016}%
{Liu \MakeLowercase{\textit{et al.}}: Query-Adaptive Hash Code Ranking for Large-Scale Multi-View Visual Search}
\IEEEcompsoctitleabstractindextext{%
\begin{abstract}
Hash based nearest neighbor search has become attractive in many applications. However, the quantization in hashing usually degenerates the discriminative power when using Hamming distance ranking. Besides, for large-scale visual search, existing hashing methods cannot directly support the efficient search over the data with multiple sources, and while the literature has shown that adaptively incorporating complementary information from diverse sources or views can significantly boost the search performance. To address the problems, this paper proposes a novel and generic approach to building multiple hash tables with multiple views and generating fine-grained ranking results at bitwise and tablewise levels. For each hash table, a query-adaptive bitwise weighting is introduced to alleviate the quantization loss by simultaneously exploiting the quality of hash functions and their complement for nearest neighbor search. From the tablewise aspect, multiple hash tables are built for different data views as a joint index, over which a query-specific rank fusion is proposed to rerank all results from the bitwise ranking by diffusing in a graph. Comprehensive experiments on image search over three well-known benchmarks show that the proposed method achieves up to 17.11\% and 20.28\% performance gains on single and multiple table search over state-of-the-art methods.
\end{abstract}

}

\maketitle

\IEEEdisplaynotcompsoctitleabstractindextext

%
\IEEEpeerreviewmaketitle

\section{Introduction}
\label{sec:intro}
Hash based nearest neighbor search has attracted great attentions in many areas, such as large-scale visual search \cite{Tao:2006:pami,He:2012a,Liu:2014:pr,Song:2014,Qin:2015,Liu:2015:tip,Zhang:2015:tmi}, object detection \cite{Dean:2013}, classification \cite{Mu:2014,Liu:2016:cvpr}, recommendation \cite{Liu:2014}, etc. Locality-Sensitive Hashing (LSH) serves as a wide paradigm of hash methods that maps the similar data into similar codes \cite{Charikar:2002}. As the most well-known method, random projection based hashing method projects and quantizes high-dimensional data to the binary hash code, with compressed storage and fast query speed by computing the Hamming distance \cite{Datar:2004}.

{Even with the sound theoretical guarantee, the random strategy usually requires a large number of hash functions to achieve desired discriminative power. Motivated by the concept of locality sensitivity, many following work devote to pursuing more compact hash codes in unsupervised \cite{Weiss:2008,Liu:2011,Gong:2011,Norouzi:2013,Shen:2013,LiuLi:2015:tnnls} and supervised manner \cite{Liu:2012:cvpr,Liu:2014:tommcap,Lin:2015} using different techniques including nonlinear hash functions \cite{Kulis:2009b,Shen:2015,Lin:2015:tpami,Liu:2016:tc}, multiple features \cite{Song:2011,Liu:2014:pr,Liu:2015:iccv,LiuLi:2015:tip}, multiple bits \cite{Kong:2012,Liu:2011,Deng:2015}, bit selection \cite{Mu:2012:ijmir,Liu:2013:cvpr}, multiple tables \cite{Liu:2013:aaai,Liu:2015:iccv,Liu:2016:tip}, discrete optimization \cite{Liu:2014:nips}, deep learning \cite{Lai:2015:cvpr}, structured quantization \cite{Gong:2013,Yu:2014}, projection bank \cite{LiuLi:2015:iccv}, online sketch \cite{Leng:2015}, etc.}

The hashing techniques, integrating the property of the subspace methods \cite{Tao:2007:pami,Tao:2009:pami} and the efficient bit manipulations, can help achieve compressed storage and fast computation in large-scale nearest neighbor search. However, there exist more than one buckets that share the same Hamming distance to the query, and subsequently samples falling in these buckets will be ranked equally according to the Hamming distance. Therefore, the quantization in hashing loses the exact ranking information among the samples, and thus degenerates the discriminative power of the Hamming distance measurement. To improve the ranking accuracy using Hamming distance, it is necessary to enable fine-grained ranking by alleviating the quantization loss.

As in many traditional applications like the classification \cite{Liu:2016:pami} and the retrieval \cite{Liu:2012:prl}, in hashing a weighting scheme can serve as one of the most powerful and successful techniques, which assesses the quality of each hash bit to improve the discriminative power of the hash code. \cite{Jiang:2011} proposed a query-adaptive Hamming distance ranking method using the learned bitwise weights for a diverse set of predefined semantic concept classes. \cite{Zhang:2012} and \cite{Zhang:2013} respectively studied a query-sensitive hash code ranking algorithms (QsRank and WhRank) based on the data-adaptive and query-sensitive bitwise weights without compressing query points to discrete hash codes. Despite of the progress with the promising performance, there are still certain limitations when using these methods. For instance, these methods are usually designed for projection based hashing algorithms rather than cluster based ones like K-means hashing \cite{He:2013}. Moreover, methods like WhRank heavily rely on the assumption of data distribution (eg., Laplace distribution for spectral hashing \cite{Weiss:2008}), which is not always true in practice.

Another promising solution to enabling fine-grained ranking is incorporating information from multiple views. {For example, in many applications, especially the large-scale search, prior studies have proved that adaptively incorporating complementary information from diverse sources or views for the same object (\eg, images can be described by different visual features like SIFT and GIST) can help boost the performance \cite{Zhang:2012:eccv,Deng:2013,Hong:2014,Liu:2014:pr,Liu:2015:tip,Hong:2016,Liu:2016:tip}.} Though there are a few research on multiple feature hashing \cite{Liu:2014:pr,Liu:2015:tip,Liu:2015:iccv} which learn discriminative hash functions and improve the search performance by adaptively leveraging multiple cues, but till now there is rare work regarding the unified solution to building multiple hash tables from complementary source views with fine-grained ranking. In practice, indexing using hash tables will be more computationally feasible, due to the fast search in a constant time using table lookup \cite{Liu:2013:aaai,Dean:2013,Liu:2016:tip}, and can significantly enhance the discriminative power of the hash codes in practice \cite{Liu:2013:aaai,Liu:2015:iccv}.

To the end, in this paper we simultaneously study both techniques and propose a query-adaptive hash code fusion based ranking method over multiple tables with multiple views. For each hash table learned from each view, we exploit the similarities between the query and database samples, and learn a set of query-adaptive bitwise weights that characterize both the discriminative power of each hash function and their complement for nearest neighbor search. Assigning different weights to individual hash bit will distinguish the results sharing the same hamming distance, and obtain a more fine-grained and accurate ranking order. Compared to existing methods, our method is more general for different types of hashing algorithms, without strict assumptions on the data distribution. Meanwhile, it can faithfully enhance the overall discriminative power of the weighted Hamming distance.

As to multi-view multiple tables, a straightforward, yet efficient solution is to respectively build hash tables from each source view and then combine them as the joint table index. Such solution can directly support any type of source and existing hashing algorithms, and be easily accomplished for general applications. To further adaptively fuse search results from multiple sources in an desired order, many studies have been proposed in the literature \cite{Zhang:2012:eccv,Deng:2013,Xu:2015:pami,Wang:2015:if,Chen:2015:tip}. These powerful rank fusion techniques faithfully boost the search performance, but in practice usually are not computationally feasible for hash table search, because they heavily depend on the fine-grained ranking. For instance, \cite{Deng:2013} and \cite{Zhang:2012:eccv} accessed the data similarities relying on the raw features and exact neighbor structures respectively, which can neither be loaded in the memory nor be dynamically updated for dynamic dataset. To adaptively fuse the query-specific ordered results without operating on the raw features, an anchor representation for candidates in each source view is introduced to characterize their neighbor structure by a neighbor graph according to their weighted Hamming distances, and then all the candidates are reranked by randomly walking on the merged multiple graphs.

To our best knowledge, this is the first work that comprehensively studies the large-scale search based on hash table indexing with multiple sources. Compared to existing methods, our method is more generic for different types of hashing algorithms and data sources, without strict assumptions on the data distribution and memory consumption. Meanwhile, it can faithfully enhance the overall ranking performance for nearest neighbor search from the aspects of fine-grained code ranking and multiple feature fusion. {Note that the whole paper extends upon a previous conference publication \cite{Ji:2014} which is mainly concentrated on hash code ranking in a single hash table generated from one view. In this paper, we conduct additional exploration on fine-grained ranking technique over multiple tables with multiple features, and provide amplified experimental results}. The remaining sections are organized as follows: The query-adaptive bitwise weighing approach is present in Section \ref{sec:qsrf}. Section \ref{sec:fusion} elaborates on the graph-based fusion of multiple table ranking. In Section \ref{sec:exp} we evaluate the proposed method with state-of-the-art methods over several large datasets. Finally, we conclude in Section \ref{sec:con}.

\begin{figure}[t]
\centering
  \vspace{-0in}
  \includegraphics[width=0.9\linewidth]{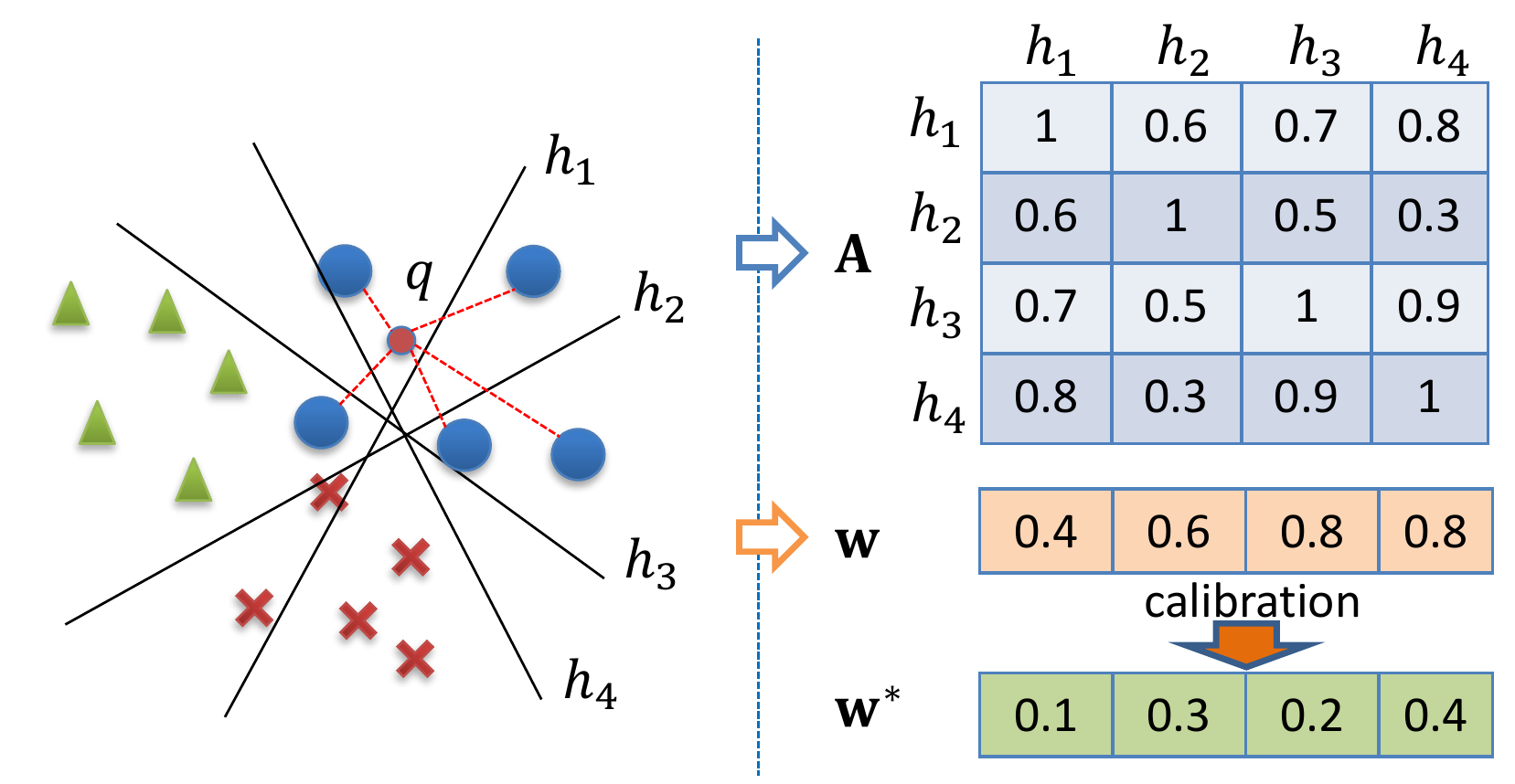}
   \vspace{-0in}
  \caption{Demonstration of the proposed query-adaptive bitwise weighting. }
  \label{fig:idea1}
    \vspace{-0in}
\end{figure}
\section{Query-Adaptive Hash Code Ranking} \label{sec:qsrf}
In this section, we will first focus on the fine-grained ranking in each hash table, and propose the query-adaptive ranking method (named QRank) using weighted hamming distance. It utilizes the similarity relationship between the query point and its neighbors in the database to measure the overall discriminative power of the hash code, simultaneously considering both the quality of hash functions and their correlations.

\subsection{Query-Adaptive Weight}\label{subsec:qaw}
Given a set of $N$ training samples $\{\mathbf{x}_i\in \mathbb{R}^D,i=1,\ldots,N\}$ and a set of $B$ hash functions $H(\cdot)=\{h_1(\cdot),\ldots,h_B(\cdot)\}$, each training sample $\mathbf{x}_i$ is encoded into hash bit $y_{ik}=h_k(\mathbf{x}_i)\in \{-1,1\}$ by the $k$-th hash function. With the hash functions and corresponding weights $\mathbf{w}$, the weighted Hamming distance between any two points $\mathbf{x}_i$ and $\mathbf{x}_j$ are usually defined as $d_h(\mathbf{x}_i,\mathbf{x}_j)=\sum_{k=1}^B w_k(y_{ik}\otimes y_{jk})$.

To improve the ranking precision of the weighted distance $d_h$, a data-dependent and query-adaptive $w_k$ should be learnt to characterize the overall quality of the $k$-th hash function in $H$. Intuitively, for a query point $\mathbf{q}$, a hash function well preserving $\mathbf{q}$'s nearest neighbors $\text{NN}(\mathbf{q})$ (eg., $h_3$ and $h_4$ in Figure \ref{fig:idea1}) should play a more important role in the weighted Hamming distance, and thus a larger weight should be assigned to it.

Formally, for a high-quality hash function $h_k$, if $\mathbf{p}\in \text{NN}(\mathbf{q})$, then the higher its similarity $s(\mathbf{p},\mathbf{q})$ to $\mathbf{q}$, the larger the probability that $h_k(\mathbf{q})=h_k(\mathbf{p})$. Therefore, based on the neighbor preservation of each hash function, we define its weight using the spectral embedding loss \cite{Weiss:2008,Liu:2013:cvpr}:
\begin{equation}\label{}
\begin{split}
    w_k &= -\frac{1}{2}\sum_{\mathbf{p}\in \text{NN}(\mathbf{q})} s(\mathbf{q},\mathbf{p})\|h_k (\mathbf{q})-h_k (\mathbf{p})\|^2\\
   & =\sum_{\mathbf{p}\in \text{NN}(\mathbf{q})} s(\mathbf{q},\mathbf{p})h_k (\mathbf{q})h_k (\mathbf{p})+const.
\end{split}
\end{equation}
where we constrain $\sum_{\mathbf{p}\in NN(\mathbf{q})} s(\mathbf{q},\mathbf{p})=1$. Note that in the above definition the similarity can be adaptively tailored for different scenarios.

To make the weight positive and sensitive to the capability of neighbor preservation, in practice we use the following form with $\gamma>0$:
\begin{equation}\label{}
w_k = \exp \left[\gamma\sum_{\mathbf{p}\in \text{NN}(\mathbf{q})} s(\mathbf{q},\mathbf{p})h_k (\mathbf{q})h_k (\mathbf{p})\right].
\end{equation}\label{}

In the above definition, one important question is how to efficiently find the query's nearest neighbors $\text{NN}(\mathbf{q})$ at the online query stage. It is infeasible to find the exact ones among the whole database \cite{Hong:2015}. One way to speedup the computation is choosing $N_l\ll N$ landmarks to represent the database using various techniques like K-means, where the approximated nearest neighbors can be quickly discovered by linear scan. We adopt the simple way by randomly sampling $N_l$ points as the landmarks at the offline training stage.

\subsection{Weight Calibration}
As Figure \ref{fig:idea1} demonstrates that hash function $h_3$ and $h_4$ shows satisfying capability of neighbor preservation for query $\mathbf{q}$, but the large correlation between them indicates undesired redundancy between them. Instead, though function $h_2$ performs worse than $h_3$ and $h_4$, but it serves as a complement to $h_4$, with which they together can well preserve all neighbor relations of $\mathbf{q}$. This observation motivates us to further calibrate the query-adaptive weights, taking the correlations among all hash functions into consideration.

Given any pair of hash functions $h_i$ and $h_j$ from $H$, if they behave similarly on a certain set of data points (i.e., the hash bits encoded by them are quite similar), then we can regard that the two hash functions are strongly correlated. In practice, to improve the overall discriminative power of the hash codes, uncorrelated (or independent) and high-quality hash functions should be given higher priority.

Since computing higher-order independence among hash functions is quite expensive, we approximately evaluate the independence based on the pair-wise correlations between them. Specifically, we introduce the mutual independence between hash functions based on the mutual information $\text{MI}({y}_i,{y}_j))$ between the bit variables $\mathbf{y}_i$ and $\mathbf{y}_j$ generated by hash function $h_i$ and $h_j$:
\begin{equation}\label{}
    a_{ij} = \exp \left[-\lambda \text{MI}(\mathbf{y}_i,\mathbf{y}_j)\right],
\end{equation}
where $\lambda > 0$ and $a_{ij} = a_{ji}$, forming a symmetrical independence matrix $\mathbf{A} = (a_{ij})$.

Then we calibrate the query-adaptive weights $w_k$ by reweighing it using a positive variable $\pi_k$. Namely, the new bitwise query-adaptive weight is given by
\begin{equation}\label{}
w_k^*=w_k\pi_k,
\end{equation}
which should overall maximize both the neighbor preservation and the independence between hash functions. We formulate it as the following quadratic programming problem:

\begin{equation}\label{eqn:prob1}
\begin{split}
\max_{\pi} \quad & \sum_{ij}w_i^{*}w_j^{*}a_{ij}\\
s.t. \quad & \mathbf{1}^{T} \pi = 1,\ \pi\succeq 0.
\end{split}
\end{equation}
The above problem can be efficiently solved by a number of powerful techniques like replicator dynamics \cite{Liu:2013:cvpr}.
%

\subsection{Data-Dependent Similarity}
As aforementioned, the similarity between the query and database samples plays an important role in pursuing query-adaptive weights in Section~\ref{subsec:qaw}. Since in practice data points are usually distributed on certain manifold, the standard Euclidean metric cannot strictly capture their global similarities. To obtain similarity measurement adaptive to different datasets, we adopt the anchor graph to represent any sample $\mathbf{x}$ by $z(\mathbf{x})$ based on their local neighbor relations to anchors points ${\mathcal{U}} = \{\mathbf{u}_k \in \mathbb{R}^{d}\}_{k=1}^{N_l}$, which can be generated efficiently by clustering or sampling \cite{Liu:2011}:
\begin{equation}\label{eqn:anchors}
\left[z(\mathbf{x})\right]_{j}=\left\{
        \begin{array}{cl}
          \frac{\mathcal{K}(\mathbf{x}, \mathbf{u}_j)}{\sum_{\mathbf{u}_{j'} \in \text{NN}(\mathbf{x})}\mathcal{K}(\mathbf{x},\mathbf{u}_{j'})}, & \text{if } \mathbf{u}_j \in \text{NN}(\mathbf{x})\\
          0, & \text{otherwise}
        \end{array}
      \right.
\end{equation}
where $\text{NN}(\mathbf{x})$ denotes ${\mathbf{x}}$'s nearest anchors in ${\mathcal{U}}$ according to the predefined kernel function $\mathcal{K}(\mathbf{x},\mathbf{u}_j)$ (e.g., Gaussian kernel). The highly sparse $z(\mathbf{x})$ serves as a nonlinear and discriminative feature representation, and can be used to efficiently approximate the data-dependent similarities between samples. Specifically, for query $\mathbf{q}$ and any point $\mathbf{p}$, their similarity can be computed by
\begin{equation}\label{eqn:sim}
    s(\mathbf{p},\mathbf{q}) = \exp(-\|z(\mathbf{p})-z(\mathbf{q})\|^2/\sigma^2),
\end{equation}
where $\sigma$ is usually set to the largest distance between $z(\mathbf{p})$ and $z(\mathbf{q})$.

\begin{figure}[t]
\centering
  \vspace{-0.0in}
  \includegraphics[width=1\linewidth]{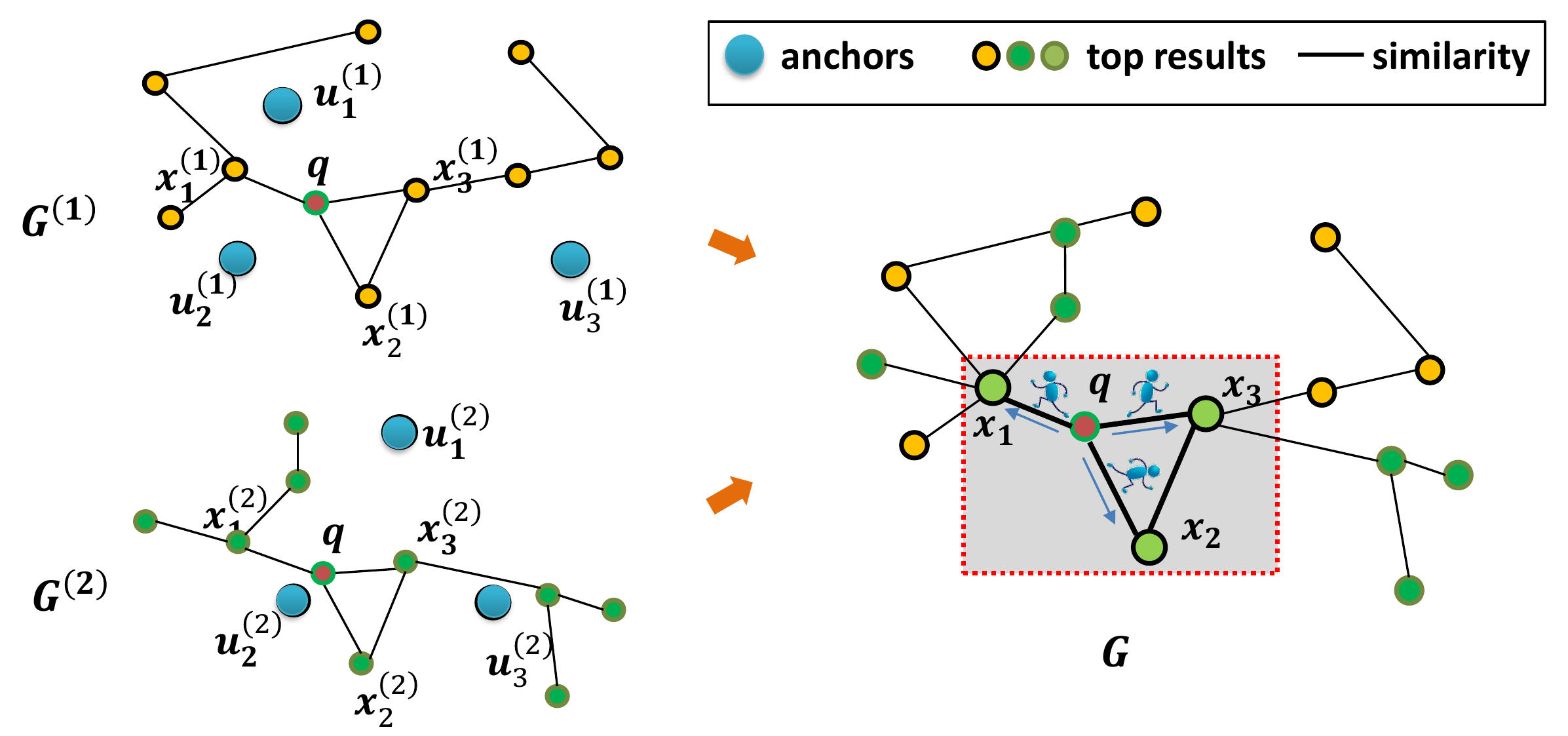}
   \vspace{-0in}
  \caption{Demonstration of the proposed table-wise rank fusion. }
  \label{fig:idea}
    \vspace{-0in}
\end{figure}

\section{Multi-View Graph-based Rank Fusion}\label{sec:fusion}
Now we turn to the fine-grained ranking over multiple hash tables with multiple view information. In this section, we will present a query specific rank fusion method that adaptively combines the ranking results of multiple tables. The weighted Hamming distance enables fine-grained ranking of each hash table, which improves the discriminative power of hash codes from the views of hash function quality and their correlations. To adaptively organize results from different tables in an desired order without any supervision or user feedbacks, we represent them as an edge-weighted graph, where the comparable similarities across different tables can be approximated efficiently online based on the neighborhood captured by anchors in the weighted Hamming space. Finally, with the consistent similarity measurement, the common candidates and their neighbors from different tables will be promoted with higher probabilities to be ranked top using a localized PageRank algorithm over the fused graph. Figure \ref{fig:idea} demonstrates the proposed rank fusion method based on probability diffusion on the merged graph.

\subsection{Graph Construction}
For the query $\mathbf{q}$, multiple candidate sets will be retrieved in hash tables built from $M$ different sources. We represent the candidate set corresponding to the $m$-th source/table, including the query, as an edge-weighted graph $\mathbf{G}^{(m)}=(\mathbf{V}^{(m)}, \mathbf{E}^{(m)}, \mathbf{\Omega}^{(m)})$. The vertices $\mathbf{V}^{(m)}$ in the graph correspond to each element in the candidate set, while the edge connections $\mathbf{E}^{(m)}$ indicate the neighbor relations, whose weights $\mathbf{\Omega}^{(m)}_{ij}$ are defined by the similarity ${\mathbf{S}}^{(m)}_{ij}$ between the candidates $\mathbf{x}_i$ and $\mathbf{x}_j$. The graph is quite powerful to characterize the local neighbor structures of the candidates along a manifold.

In practice, similar candidates usually own a large portion of common neighbors. Therefore, it is critical to capture the local neighbor structure centered at each candidate point in the similarity measurement. In traditional methods the similarity measurement usually depends on the raw features or neighbor structures process beforehand, \eg, using the Euclidean distance between raw features \cite{Deng:2013} or Jaccard index (common neighbors) in the neighborhood \cite{Zhang:2012:eccv}. Both solutions are infeasible for large-scale datasets, due to the explosive memory consumption and expensive updating.

To avoid this bottleneck in the online graph construction, we estimate the neighbor relations by employing the anchor approximation in the weighted Hamming space. A small set of $K$ anchors are selected as the prototypes to together locate each data point in the specific space. Therefore, each point can be characterized by its nearest anchors, and thus their neighbor relations can be determined by checking whether sharing similar nearest anchors.

Formally, for candidate set $\mathbf{V}^{(m)}$ of $m$-th source, the $K$ ($K \ll N$) anchor points ${\mathcal{U}}^{(m)} = \{\mathbf{u}_k^{(m)} \in \mathbb{R}^{D_m}\}_{k=1}^K$ are selected from the database to characterize the inherent neighbor structures. Thus, any data point $\mathbf{x}_i$ can be represented by its nearest anchors in a vector form, whose elements are denoted by the truncated similarities $\mathbf{Z}^{(m)}_{ij}$ with respect to the $m$-th source:
\begin{equation}\label{eqn:anchors}
\mathbf{Z}_{ij}^{(m)}=\left\{
        \begin{array}{cl}
          \frac{\exp\left(-{d_H(\mathbf{x}^{(m)}_i, \mathbf{u}^{(m)}_j)}/{\sigma_H}\right)}{\sum_{j' \in \langle i \rangle^{(m)}}\exp\left(-{d_H(\mathbf{x}^{(m)}_i, \mathbf{x}^{(m)}_{j'})}/{\sigma_H}\right)}, & \text{if } j\in \langle i \rangle^{(m)}\\
          0, & \text{otherwise}
        \end{array}
      \right.
\end{equation}
where $\langle i \rangle^{(m)} \subset [1:K]$ denotes the indices of $s$ $(s \ll K)$ nearest anchors of point $\mathbf{x}_i$ in ${\mathcal{U}}^{(m)}$ according to the weighted Hamming distance, and $\sigma_H$ is set to the maximum Hamming distance.

The symmetric similarity ${\mathbf{S}}^{(m)}$ between candidates in $\mathbf{V}^{(m)}$ can be defined by the inner product between their anchor representations in the $m$-th feature space:
\begin{equation}\label{eqn:similarity}
{\mathbf{S}}^{(m)}_{ij} = \frac{1}{\lambda_i^{(m)}} \mathbf{Z}^{(m)}_{i\cdot}\mathbf{Z}_{j\cdot}^{(m)T} + \frac{1}{\lambda_j^{(m)}} \mathbf{Z}^{(m)}_{j\cdot}\mathbf{Z}_{i\cdot}^{(m)T}
\end{equation}
with $\lambda_i^{(m)} = \sum_{j} \mathbf{Z}^{(m)}_{i\cdot}\mathbf{Z}_{j\cdot}^{(m)T}$. Since $\mathbf{Z}^{(m)}\in \mathbb{R}^{N\times K}$ is highly sparse with $s$ nonzero elements in each row summing to 1, the approximated similarity $\mathbf{S}^{(m)}$ will be also quite sparse, where only those data sharing same nearest anchors will be regarded as neighbors with non-zero similarities.

\begin{algorithm}[tp!]
\caption{\textbf{Q}uery-\textbf{s}pecific \textbf{R}ank \textbf{F}usion over Hash Tables with Multiple Sources (QsRF)} \label{alg:qq}
\begin{algorithmic}
\STATE // \emph{offline stage, building hash tables}
\REQUIRE data $\{\mathbf{x}_i, i = 1,\ldots,N\}$ of multiple sources, hashing algorithm $\mathcal{F}$
\FOR{$m$-th source}
\STATE generate anchors ${\mathcal{U}}^{(m)}$;
\STATE generate functions $H =\{h_1(\cdot),\ldots,h_B(\cdot)\}$ using $\mathcal{F}$;
\STATE encode each source of $\mathbf{x}_i$ to $y_{ik}=h_k(\mathbf{x}_i)$;
\STATE compute independence $a_{ij}$ and build hash table $\mathcal{T}_m$;
\ENDFOR
\STATE // \emph{online stage, searching hash tables}
\REQUIRE query $\mathbf{q}$, hash tables $\{\mathcal{T}_m\}_{m=1}^M$, independence $a_{ij}$;
\FOR{$m$-th table $\mathcal{T}_m$}
\STATE compute the query-specific bitwise weights $w^*_k$;
\STATE retrieve the top ranked data $\mathbf{V}^{(m)}$ according to $d_H$;
\STATE approximate neighbor similarity $\mathbf{\Omega}^{(m)}$ using anchors;
\STATE build graph $\mathbf{G}^{(m)}=(\mathbf{V}^{(m)}, \mathbf{E}^{(m)}, \mathbf{\Omega}^{(m)})$;
\ENDFOR
\STATE fuse graph $\mathbf{G}^{(m)}$ as graph $\mathbf{G} = (\mathbf{V},\mathbf{E},\mathbf{\Omega})$;
\STATE compute rank score $\mathbf{r}^*$ and reorder all candidates in $\mathbf{V}$.
\end{algorithmic}
\end{algorithm}

\subsection{Rank Fusion}
After obtaining multiple graphs $\mathbf{G}^{(m)} = (\mathbf{V}^{(m)},\mathbf{E}^{(m)},\mathbf{\Omega}^{(m)})$ from different hash tables corresponding to each type of source, we fuse them together as one graph $\mathbf{G} = (\mathbf{V},\mathbf{E},\mathbf{\Omega})$, where the vertices correspond to all candidates without repeats. Subsequently, for each pair of candidates if  there is an edge between them in any $\mathbf{G}^{(m)}$, there will be a connection between them in $\mathbf{G}$, and its weight will be the superposition of all weights in $\mathbf{G}^{(m)}$. This will guarantee that the common candidates and their neighbors from different hash tables show high homogeneity, and thus have priorities to be ranked top. Note that the anchor-based representation not only captures the local neighbor structure along the manifold, but also makes the similarity comparable across different rank lists after normalization. Therefore, without any prior, we regard multiple retrieval results equally and simply sum up their edge weights. Formally,
\begin{equation}\label{eqn:fuse}
\begin{array}{lll}
\mathbf{V} & = &\bigcup_{m} \mathbf{V}^{(m)},\\
\mathbf{E} & = &\bigcup_{m} \mathbf{E}^{(m)}, \\
\mathbf{\Omega}_{i,j} & = &\sum_{m}\mathbf{\Omega}^{(m)}_{i,j}=\sum_{m}{\mathbf{S}}^{(m)}_{ij}.
\end{array}
\end{equation}

For multiple table search with different sources, the strong connected candidates in $\mathbf{G}$ show great and consistent visual similarities in different views, which implies an inherent neighbor relations between them, forming a dense subgraph. To discover the most relevant candidates for the query, this motivates us to perform connectivity analysis on the graph, where the desired ones will have high probability to be visited from the query.

One popular and powerful technique is random walk, interpreted as a diffusion process \cite{Zhang:2012:eccv} and succeeding in the Google PageRank system. Specifically, the walk in $\mathbf{G}$ will sometimes (with a small probability $1-\alpha$, where empirically $\alpha > 0.8$) jump to a potentially random result with a fixed distribution $\mathbf{r}$, or transit in a random walk way according to the vertex connection matrix $\mathbf{P} = \mathbf{\Omega}$. For query-specific ranking, distribution $\mathbf{r}$ for query $\mathbf{q}$ will be empirically set to a large probability (\eg, 0.99), while others being uniform, satisfying $\mathbf{r}^T \mathbf{1}=1$. Formally the visiting probability $\mathbf{r}$ of each vertex after one-step random walk from the $t$-th step will be updated as follows:
\begin{equation}\label{eqn:p}
\mathbf{r}^{(t+1)} = (1-\alpha)\pi + \alpha \mathbf{P}^T \mathbf{r}^{(t)}.
\end{equation}
With a number of random walks, the visiting probability $\mathbf{r}$ of each candidate will converge to an equilibrium state, where a higher probability reflects a higher relevance to the query.

Using the above updating strategy, the equilibrium distribution can be obtained in a close form, \ie,  $\mathbf{r}^* = (1-\alpha)(\mathbf{I}-\alpha \mathbf{P}^T)^{-1}\pi$. Note that such solution requires matrix inverse, which will bring expensive computation when a large number of candidates participate in the rank fusion. To speedup this process, usually the iterative updating in (\ref{eqn:p}) is preferred in practice. Algorithm \ref{alg:qq} list the detailed procedures at both offline and online stages.

\subsection{Online Query}
At the offline stage, for each view we build a hash table using certain hashing algorithms. Both $N_l$ sampled landmarks and $K$ anchors can be obtained efficiently, and the landmarks can be represented using anchors in $O(N_lK)$. At the online stage, for each table the query in the corresponding feature representation is transformed into anchor representation and used to compute the query's similarities to the landmarks in $O(N_l)$, and then the query-adaptive weights can be obtained by quadratic programming (\ref{eqn:prob1}) in polynomial time, and the results are fast ranked according to the weighted hamming distances. With the ranked result sets from different tables, multiple graph can be built over all the candidates which are empirically limited to the top $N_k \ll N$ (\eg, 1,000 in our experiments) ones in practice. Subsequently, the graph construction based on the anchor representation will cost $O(MN_k K + N_k^2K)$ time, which is much less than the database size $N$. Finally, the reranking process fusing multiple ranking lists can be completed in a very few  random walk iterations of $O(N_k^2)$ complexity. On the whole, the proposed rank fusion method can support fast online search in a sublinear time with respect to the database size.

\section{Experiments}\label{sec:exp}
In this section, we comprehensively evaluate the effectiveness of our proposed method consisting of query-adaptive hash code ranking (QRank) and query specific rank fusion (QsRF) method, over several real-world image datasets.

\subsection{Data Sets}
To evaluate the proposed methods, we conduct extensive experiments on the real-world image datasets as follows:

\textbf{MNIST} includes 70K 784 dimensional images, each of which is associated with a digit label from `0' to `9'. Each example is a 28$\times$28 image and is represented by a 784 dimensional vector.

\textbf{CIFAR-10} contains 60K $32 \times 32$ color images of 10 classes and 6K images in each class. For each image, we extract 300-D bag-of-words (BoW) quantized from dense SIFT features and 384-D GIST feature.

 \textbf{TRECVID} \cite{Yu:2012} is a large-scale image dataset built from the TRECVID 2011 Semantic Indexing annotation
 set with 126 fully labeled concepts, from which we select 25 most-frequent concepts. For each image, we extract 512-D GIST feature and 1000-D spatial pyramid bag-of-words feature.

\begin{figure*}[!tp]
\centering
\includegraphics[width=1\linewidth]{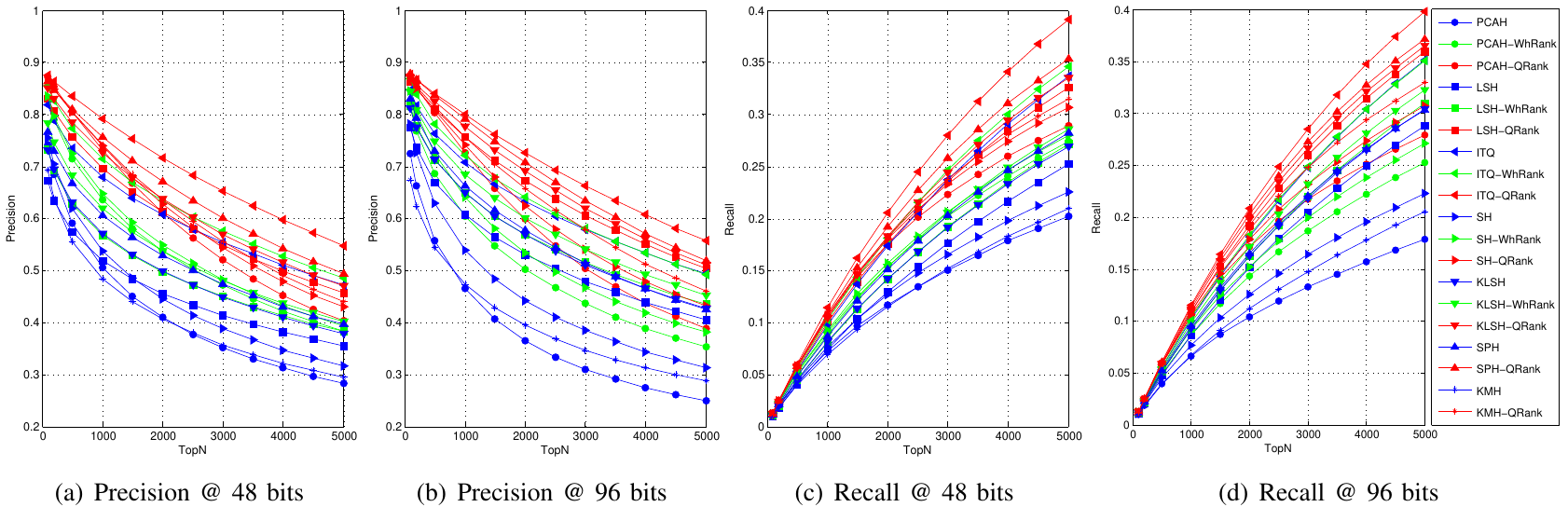}
  \caption{\small Performances comparison of different hash code ranking methods on MNIST.}
  \label{fig:pr_mnist}
\end{figure*}

\begin{figure*}[!tp]
\centering
\includegraphics[width=1\linewidth]{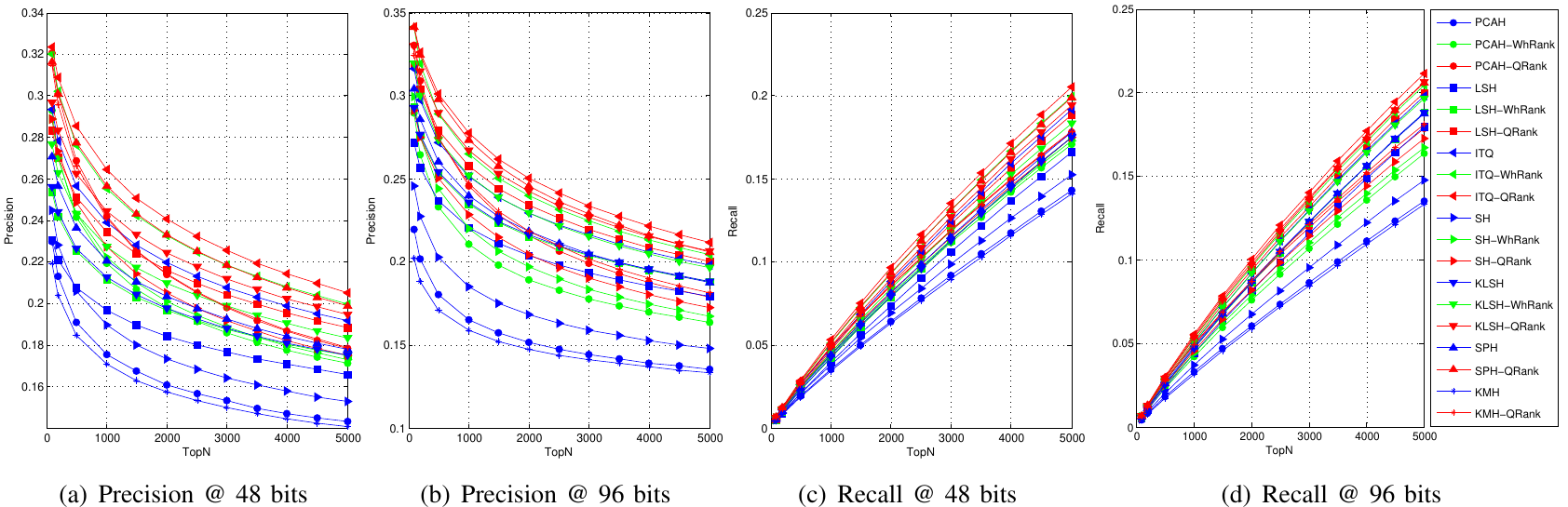}
  \caption{\small Performances comparison of different hash code ranking methods on CIFAR-10.}
  \label{fig:pr_cifar}
\end{figure*}

\begin{figure*}[!tp]
\centering
\includegraphics[width=1\linewidth]{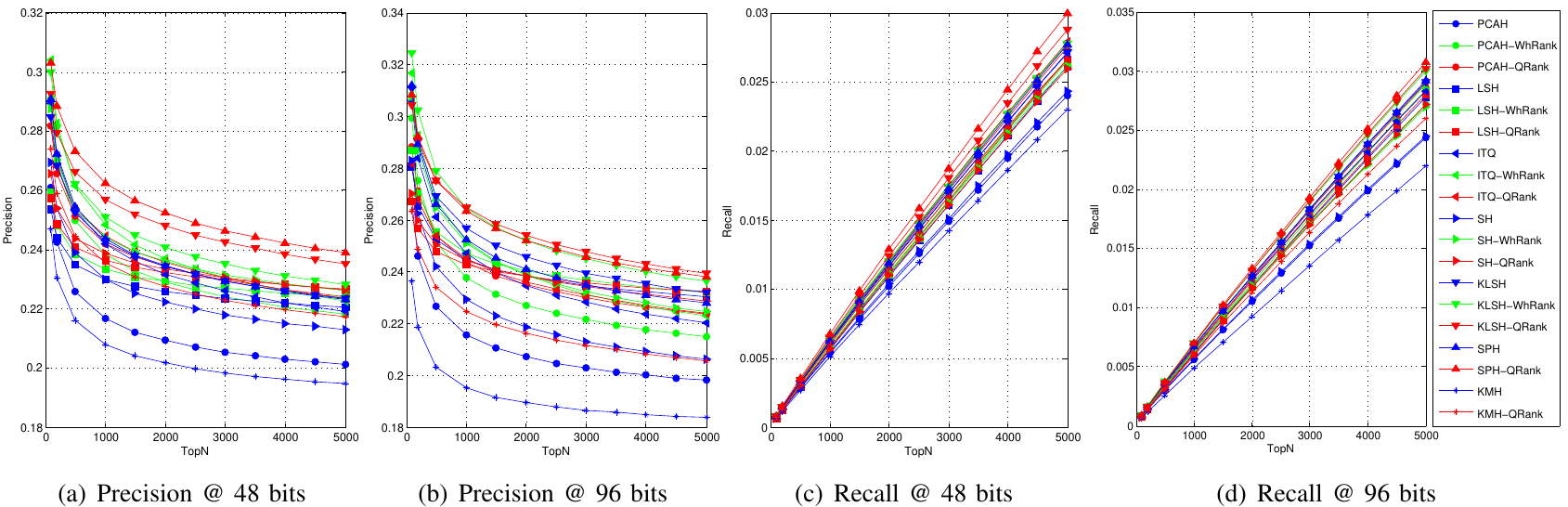}
  \caption{\small Performances comparison of different hash code ranking methods on TRECVID.}
  \label{fig:pr_trecvid}
\end{figure*}

\subsubsection{Protocols}
In the literature, there are very few studies regarding the Hamming distance ranking besides WhRank \cite{Zhang:2013} and QsRank \cite{Zhang:2012}. Since WhRank outperformed QsRank \cite{Zhang:2012} significantly as reported in \cite{Zhang:2013}, we only focus the comparison between our method and WhRank \cite{Zhang:2013} over a variety of baseline hashing algorithms including projection based and cluster based ones, which cover most types of the state-of-the-art hashing research including linear/nonlinear hashing and random/optimized hashing.

Specifically, In our experiments each hash table is constructed using hash functions generated by basic hashing algorithms. State-of-the-art hashing algorithms, including Locality Sensitive Hashing (LSH) \cite{Datar:2004}, Spectral Hashing (SH) \cite{Weiss:2008}, Kernelized LSH (KLSH) \cite{Kulis:2009b}, PCA-based Hashing (PCAH), Iterative Quantization (ITQ) \cite{Gong:2011}, Spherical Hashing (SPH) \cite{Heo:2012}, and K-means Hashing (KMH) \cite{He:2013}, are involved in generating hash functions. Among these methods, LSH, SH, KLSH, PCAH and ITQ adopted the projection paradigm to generate hash bits, while SPH and KMH exploit the cluster structure in the data. Meanwhile, LSH, PCAH and ITQ are linear methods, and SH, KLSH, SPH and KMH are nonlinear ones.

The common search method named Hamming distance ranking will be adopted in the evaluation, which ranks all points in the database according to their (weighted) Hamming distances from the query. We adopt the popular performance metric including precision and recall in our experiments. All results are averaged over 10 independent runs to suppress the randomness. {For each run, we randomly sample 5,000 images as the training data, 3,000 as query images, and 300 as the anchors for both query-adaptive weighting and the similarity approximation in each view.} The true nearest neighbors are defined as the images sharing at least one common tag.

In our experiments, we first investigate the performance of the proposed query-adaptive hash code ranking using a single feature type in one hash table. On MNIST we employ the grayscale pixel intensities of each image as the feature, and on CIFAR-10 and TRECVID we use GIST descriptors for evaluation. As to the rank fusion over multiple hash tables with multiple sources, for simplicity and similar to prior multiple feature work \cite{Song:2011,Liu:2014:pr}, we adopt two types of visual features for each set (see the dataset description above) to verify the efficiency of our proposed method. Namely, without loss of generality, we evaluate our method on image search over \textbf{CIFAR-10} (60K) and \textbf{TRECVID} (260K), and each source corresponds to one type of visual features.

\subsection{Evaluation on query-adaptive hash code ranking}
\begin{figure*}[t]
\centering
\includegraphics[width=1\linewidth]{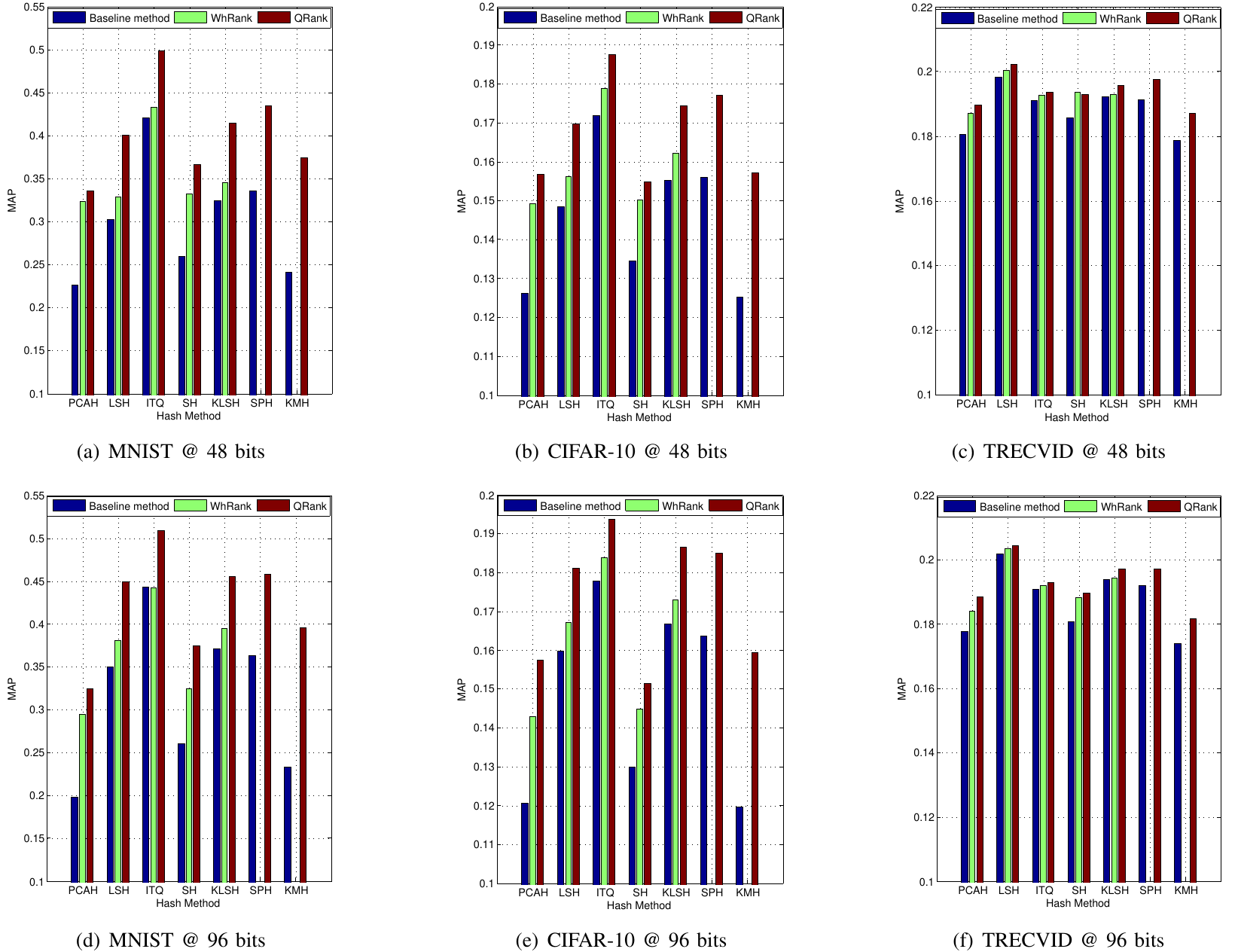}
  \caption{\small Average precision comparison on MNIST, CIFAR-10, and TRECVID.}
  \label{fig:ap_all}
\end{figure*}

Figure \ref{fig:pr_mnist} shows the precision and recall curves respectively using 48 and 96 bits on MINST. We can easily find out that both hash bit weighting methods (WhRank and QRank) achieved better performances than the baseline hashing algorithms. The observation indicates that the hash code ranking based on weighted Hamming distance hopefully serves as a promising solution to boosting the performance of hash-based retrieval. In all cases, our proposed QRank consistently achieves the superior performances to WhRank over different hashing algorithms. Figure \ref{fig:ap_all}(a) and (d) depict the overall performance evaluation using mean average precision (MAP) with 48 and 96 hash bits, where we get a similar observation that QRank outperforms WhRank significantly, e.g., compared to WhRank, QRank obtains 17.11\%, 11.36\%, 14.06\% and 11.39\% performance gains respectively over LSH, ITQ, SH and KLSH.

\begin{table}[t]
\centering{
\caption{MAP (\%) of different parts of QRank on MNIST}\label{tab:map}
\begin{tabular}{c c c c c} \hline
 & LSH & SH & PCAH & ITQ \\ \hline
Baseline & 35.53 & 25.91 & 19.87 & 44.14\\
QRank$^-$	& 40.71 & 31.39 & 22.07 & 46.87\\
QRank	&  \textbf{44.77} & \textbf{37.02}  & \textbf{32.32} & \textbf{49.15}\\\hline
\end{tabular}
} \vspace{-0in}
\end{table}
Besides MNIST, Figure \ref{fig:pr_cifar} and \ref{fig:pr_trecvid} present the recall and precision curves on CIFAR-10 and TRECVID using 48 and 96 bits. By comparing the performance of QRank with that of WhRank and baseline hashing algorithms, we get a similar conclusion as on MNIST that both QRank and WhRank can enhance the discriminative power of hash functions by weighting them elegantly, and meanwhile in all cases QRank clearly outperforms WhRank owing to its comprehensive capability of distinguishing high-quality functions. Similar conclusion can be obtained from the MAP performance comparison as the overall evaluation in Figure \ref{fig:ap_all} (b), (c), (e) and (f) with respect to different number of hash bit per table.

It is worth noting that since WhRank depends on the distribution of the projected samples, it cannot be directly applied to hashing algorithms like SPH and KMH. Instead, the proposed method derives the bitwise weight only relying on the data-dependent similarity. Therefore, it can not only capture the neighbor relations of the query, but also possess universality suiting for all hashing algorithms. Figure \ref{fig:pr_mnist} - \ref{fig:ap_all} show that QRank obtains significant performance gains over SPH and KMH in all cases.

We investigate the effect of different parts of QRank in Table \ref{tab:map} by comparing the performance of QRank with or without the weight calibration (named QRank$^-$). The results are reported over LSH, SH, PCAH and ITQ respectively using 96 bits on MNIST. It is obvious that QRank$^-$ only using query-adaptive weights without considering the redundance among hash functions is able to boost the hashing performance, but QRank appended with the weight calibration can further bring significant (up to 46.44\%) performance gains. This indicates that both the individual quality of each hash function and their complementarity are critical for fine-grained neighbor ranking. Finally, we also compare the time cost of different query adaptive ranking strategies. Though our QRank spends a little more time than WhRank on learning query-adaptive weight, on the whole this part is relatively small and the online query is quite efficient in practice (see Table \ref{tab:time}).

\subsection{Evaluation on query specific rank fusion}
\begin{table*}[tp]
\caption{AP (\%) of hash table search using different ranking methods over multiple tables on CIFAR-10 and TRECVID.}
\vspace{-0in}
\hspace{-0.3in}
\label{tab:ap}
\begin{center}
{
\begin{sc}
\begin{tabular}{ l  c  c  c   c  c  c  c  c  c  c  c}
\hline
\multirow{2}{*}{\rotatebox{0}{}} & hashing $\rightarrow$& \multicolumn{2}{c}{LSH} & \multicolumn{2}{c}{KLSH} & \multicolumn{2}{c }{PCAH} & \multicolumn{2}{c }{ITQ} & \multicolumn{2}{c }{SPH}\\\cline{2-12}

  &  methods    & AP@5  &   AP@10   & AP@5  &   AP@10    & AP@5  &   AP@10     & AP@5  &   AP@10   & AP@5  &   AP@10    \\\hline
\multirow{7}{*}{\rotatebox{90}{CIFAR-10}}
&	 F1 	&	14.93	&	15.19	&	16.09	&	16.13	&	18.38	&	18.58	&	17.94	&	 17.99	&	 18.96	&	 18.44	\\

&	 F2 	&	18.17	&	17.91	&	21.25	&	20.62	&	26.72	&	25.80	&	27.92	&	 27.55	&	 23.58	&	 22.93	\\

&	 F1-Qs 	&	15.21	&	15.39	&	16.57	&	16.44	&	19.24	&	18.70	&	18.32	&	 18.57	&	 19.52	&	 18.41	\\

&	 F2-Qs 	&	19.31	&	19.53	&	21.95	&	21.90	&	28.20	&	28.17	&	28.52	&	 28.22	&	 24.30	&	 24.34	\\

&	 MFH 	&	15.56	&	18.89	&	15.56	&	18.89	&	15.56	&	18.89	&	15.56	&	 18.89	&	 15.56	&	 18.89	\\

&	 HRF 	&	13.33	&	16.67	&	13.33	&	16.67	&	13.33	&	13.33	&	13.33	&	 17.78	&	 15.56	&	 17.78	\\

&	 F1-F2-QsRF 	&	\textbf{20.81}	&	\textbf{20.49}	&	\textbf{23.89}	&	\textbf{23.25}	&	 \textbf{29.68}	&	 \textbf{29.40}	&	\textbf{28.72}	&	\textbf{28.51}	&	\textbf{26.58}	&	 \textbf{26.50}	\\\hline

\multirow{7}{*}{\rotatebox{90}{Trecvid}}
&	 F1 	&	23.84	&	23.12	&	44.01	&	35.96	&	46.53	&	37.37	&	43.91	&	 37.09	&	 46.25	&	 38.02	\\

&	 F2 	&	32.23	&	29.34	&	42.25	&	35.52	&	46.60	&	38.01	&	38.14	&	 34.07	&	 44.31	&	 37.12	\\

&	 F1-Qs 	&	23.88	&	23.15	&	43.76	&	35.81	&	45.66	&	36.68	&	43.43	&	 36.81	&	 46.31	&	 38.11	\\

&	 F2-Qs 	&	32.24	&	29.37	&	42.34	&	35.80	&	45.94	&	37.18	&	37.88	&	 34.00	&	 44.04	&	 36.82	\\

&	 MFH 	&	22.22	&	21.11	&	22.22	&	21.11	&	22.22	&	21.11	&	22.22	&	 21.11	&	 22.22	&	 21.11	\\

&	 HRF 	&	24.44	&	23.33	&	35.56	&	34.44	&	46.67	&	33.33	&	26.67	&	 26.67	&	 33.33	&	 30.00	\\

&	 F1-F2-QsRF 	&	\textbf{38.78}	&	\textbf{33.31}	&	\textbf{48.32}	&	\textbf{40.23}	&	 \textbf{50.31}	&	 \textbf{41.41}	&	\textbf{45.92}	&	\textbf{39.86}	&	\textbf{50.30}	&	 \textbf{42.27}\\
\hline	
\end{tabular}
\end{sc}
}  \vspace{-0in}
\end{center}
\end{table*}
In this part, we further evaluate our query specific rank fusion method over several hashing algorithms. Note that our method serves as a generic solution to build multiple tables using various sources, which is compatible with most types of hashing algorithms.  To illustrate the efficiency of the proposed method (QsRF), we will compare it on several datasets with several baseline methods: the basic hashing algorithms that build a single table using information from one source based on the standard Hamming distance (F1/F2), query-specific ranking method over each hashing algorithm without rank fusion (F1/F2-Qs), multiple feature hashing algorithm (MFH) \cite{Song:2011}, and the very recent rank fusion method (HRF) that reranks candidates based on the minimal Hamming distances, which shows promising performance in multiple table search \cite{Liu:2013:cvpr,Liu:2016:tip}.

In our experiments, a specific number of hash functions are first generated using different hashing algorithms. Then all methods except MFH build hash tables using these functions in different manners. MFH will learn its hash functions and then build tables by simultaneously considering information from multiple sources.

Table \ref{tab:ap} shows the averaged precision (AP) with respect to different cutting points (@5 and @10) respectively on CIFAR-10 and TRECVID. We can easily find out that in most cases, query specific ranking (F1/F2-Qs) achieved better performances than the baseline hashing algorithms (F1/F2). The observation indicates that the hash code ranking based on weighted Hamming distance hopefully serves as a promising solution to boost the performance of hash table search. {Furthermore, in all cases our proposed QsRF consistently achieves the superior performances to all baselines over different hashing algorithms, \eg, on TRECVID, it obtains 20.28\%, 9.80\% and 8.62\% AP@5 performance gains over the best competitors when using LSH, KLSH and SPH. This is because that our graph based rank fusion can faithfully exploit the complementarity between different features and thus discover more true semantic neighbors. Besides, note that on LSH we can obtain more significant performance improvement than other learning based hashing algorithms that own better discriminative power, which indicates that the proposed method can largely compensate the quantization loss of the basic hashing algorithms.}

\begin{figure}[t]
\centering
\includegraphics[width=.9\linewidth]{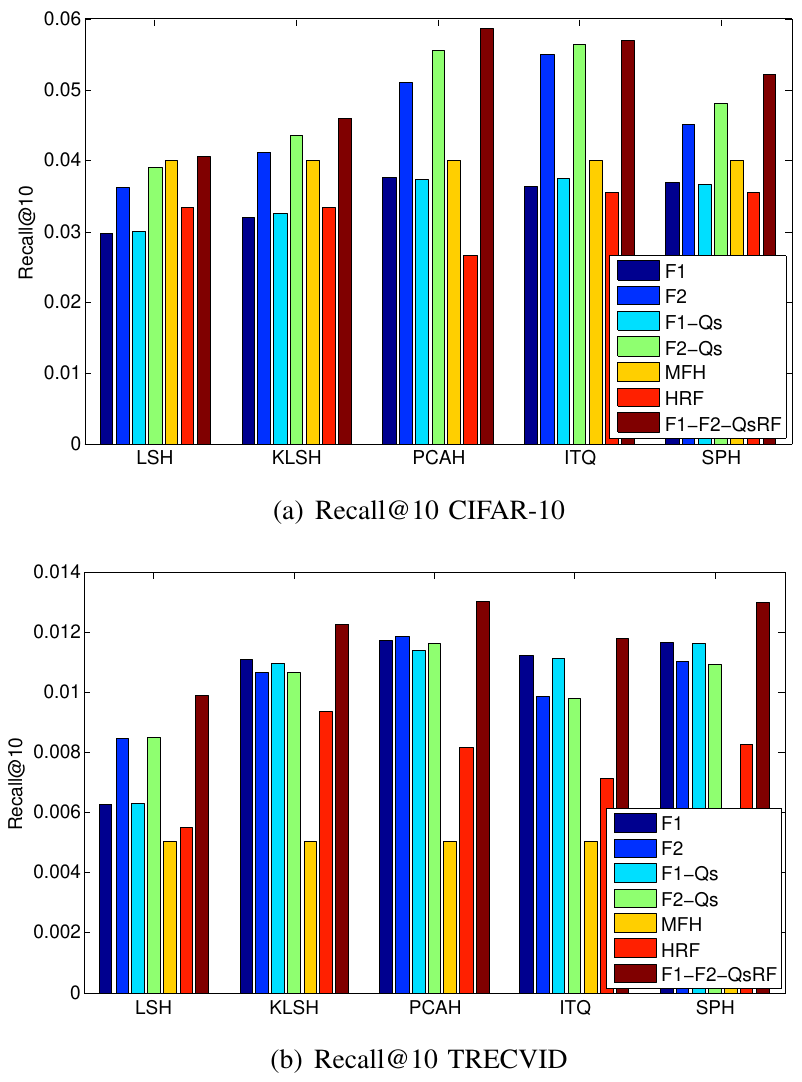}
  \vspace{-0.in}
  \caption{\small Recall performances of different ranking methods over multiple tables on CIFAR-10 and TRECVID.}
  \label{fig:r}
\end{figure}

Besides averaged precision, recall is also an important performance metric for nearest neighbor search. Figure \ref{fig:r} presents the recall rate of top 10 results using different methods on both datasets. By observing the performance of methods with query specific scheme, we get a similar conclusion that the fine-grained weighted Hamming distance can enhance the discriminative power of hash functions. {Meanwhile, QsRF can further adaptively capture and fuse the local neighbor structures in different views, and subsequently outperforms other ranking methods in all cases, with much more true relevant images returned after rank fusion. Even though the semantic complexity is quite different on CIFAR-10 and TRECVID, our method can consistently capture the intrinsic relations and obtain the best performance in most cases.}

It is worth noting that our rank fusion framework is generic for any type of binary hashing algorithms including the linear (LSH, PCAH, ITQ, etc.), nonlinear (KLSH, SPH, etc.), random (LSH, KLSH, etc.) and optimized (PCAH, ITQ, SPH, etc.) hashing. Therefore, it can be easily applied to state-of-the-art hashing research to support effective hash table search using multiple sources, achieving significant performance gains without too much efforts.

Finally, we investigate the efficiency of different methods in Table \ref{tab:time}. For multiple hash tables, the search process can be completed in parallel, which will be close to that of the single table search. Due to online computation of the query adaptive bit weights, query-specific ranking will cost a little more than standard Hamming distance ranking. For QsRF, since the candidate graph is quite sparse, the convergence of the equilibrium state distribution is computationally efficient ($O(|E|)$ time complexity), especially when adopting parallel computing and fast iterative updating. Therefore, QsRF only spends slightly more time than baseline methods. In practice, since usually only tens of hash tables are employed, it is very beneficial that QsRF can get significant performance gains, and meanwhile still guarantee fast online search.

\begin{table}[!t]
\centering{
\caption{Search time (ms) of different methods on CIFAR-10}\label{tab:time}
\begin{tabular}{c c c c c c} \hline
 & F1/F2 & F1/F2-Qs &  MFH & HRF & QsRF \\ \hline
distance rank	         & 26    &   57	& 	38  &   26  &    56 \\
rank fusion         & 0    &   0    &   0  &   21  &    36\\
total	           & 26	   &   57	& 	38  &   47  &  92\\ \hline
\end{tabular}
}
\end{table}


\section{Conclusions}\label{sec:con}
As described in this paper, we proposed a novel method for fine-grained ranking over multiple hash tables incorporating information from multiple views. At the bitwise level, the hash code ranking method named QRank learns a query-adaptive bitwise weights by simultaneously considering both the individual quality of each hash function and their complement for nearest neighbor search. At the tablewise level, a query specific rank fusion method represents results of each table using QRank as an edge-weighted graph with comparable similarities across different tables, and promotes the common candidates and their neighbors from different tables using a localized PageRank algorithm over the merged graph. Compared to state-of-the-art weighting methods, our rank fusion method serves as a general solution to enabling fine-grained ranking over multiple hash tables with multiple sources, which supports different kinds of hashing algorithms and different feature types. Due to both the query-adaptive ranking and data-dependent fusion, our method significantly and efficiently boosts the ranking performance in practice. Further work can be concentrated on learning a more discriminative similarity metric and building each hash table from multiple views with the query specific ranking.

\section{Acknowledgments}
\label{sec:ack}
This work is supported by the National Natural Science Foundation of China (61402026 and 61572388), Australian Research Council Projects (FT-130101457, DP-140102164 and LE-140100061), the Foundation of the State Key Laboratory of Software Development Environment (SKLSDE-2016ZX-04) and the Innovation Foundation of BUAA for PhD Graduates.

\bibliographystyle{IEEEtran}
\bibliography{referene}

\begin{biography}[{\includegraphics[width=1in,height=1.25in,clip,keepaspectratio]{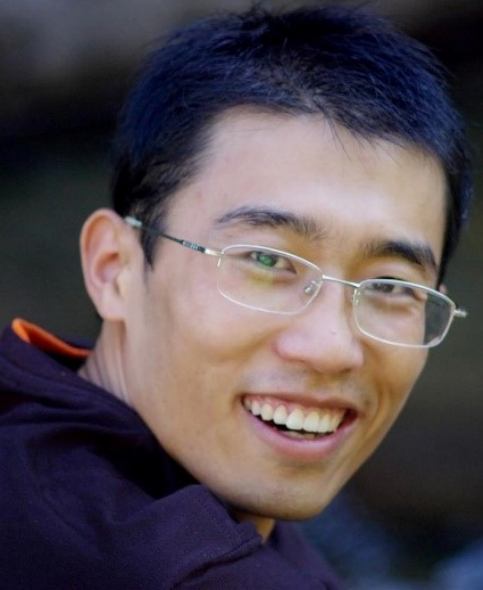}}]{Xianglong Liu}
received the BS and Ph.D degrees in computer science from Beihang University, Beijing, China, in 2008 and 2014. From 2011 to 2012, he visited the Digital Video and Multimedia (DVMM) Lab, Columbia University as a joint Ph.D student. He is currently an Associate Professor with the School of Computer Science and Engineering, Beihang University. He has published over 30 research papers at top venues like the IEEE TRANSACTIONS ON IMAGE PROCESSING, the IEEE TRANSACTIONS ON CYBERNETICS, the Conference on Computer Vision and Pattern Recognition, the International Conference on Computer Vision, and the Association for the Advancement of Artificial Intelligence. His research interests include machine learning, computer vision and multimedia information retrieval.
\end{biography}

\begin{biography}[{\includegraphics[width=1in,height=1.25in,clip,keepaspectratio]{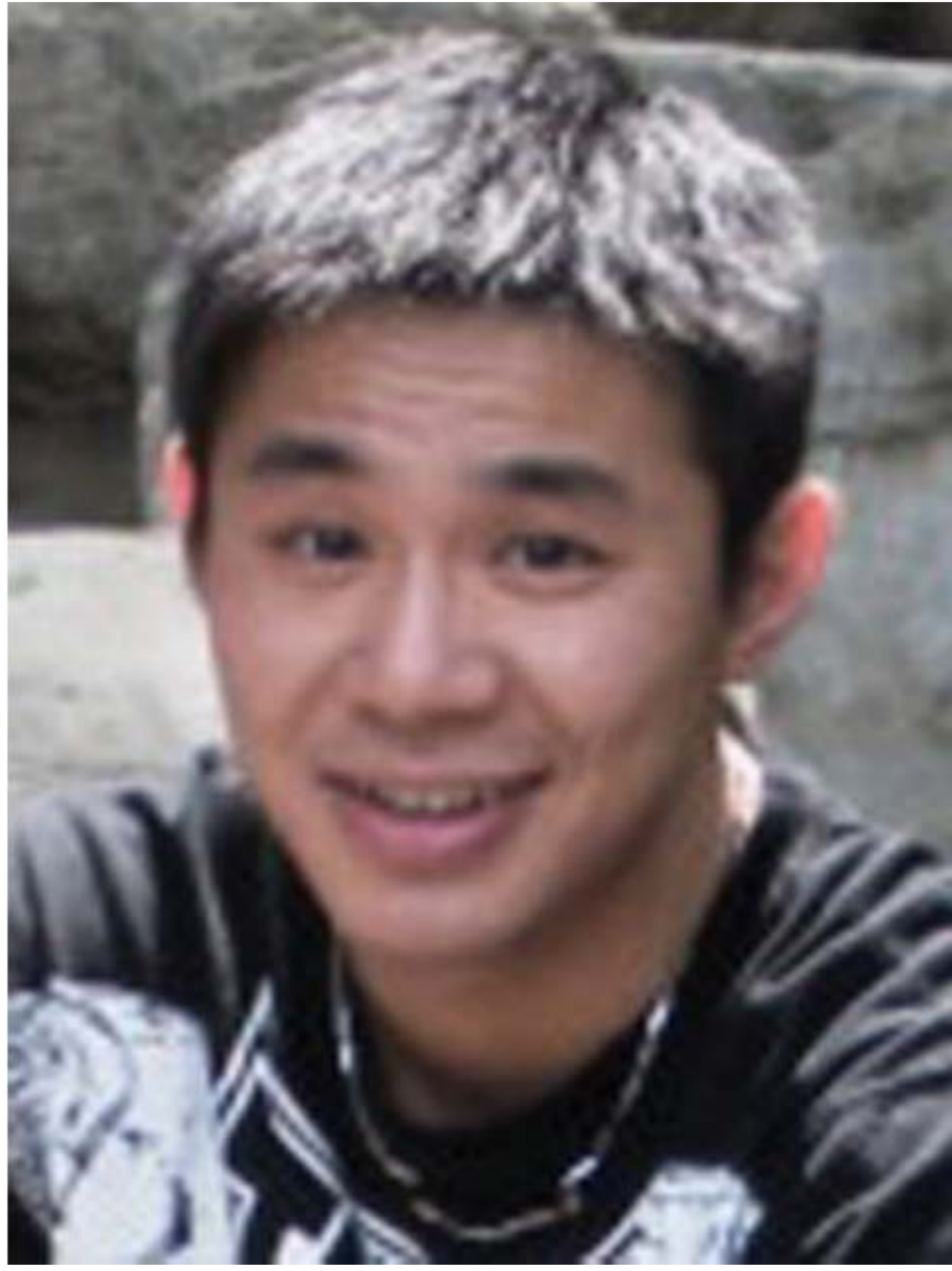}}]{Lei Huang} received the B.S. degree in computer
science from Beihang University, Beijing, China, in 2010. He is currently a Ph.D. candidate at Beihang University and is visiting Artificial Intelligence Lab at University of Michigan as a joint Ph.D student. His research interests include deep learning, semi-supervised learning, active learning,  large scale data annotation and retrieval.
\end{biography}

\begin{biography}[{\includegraphics[width=1in,height=1.25in,clip,keepaspectratio]{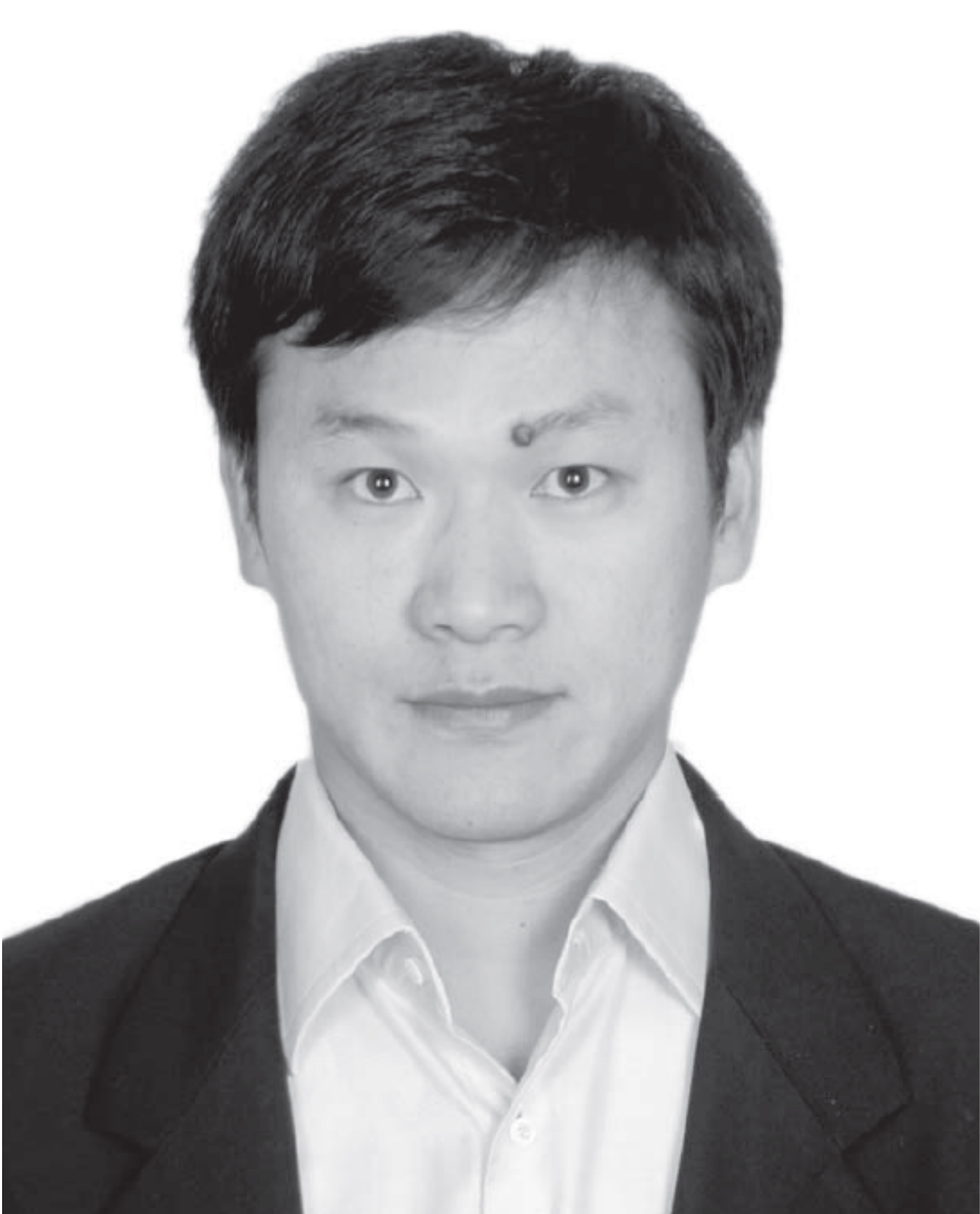}}]{Cheng Deng}
received the B.Sc., M.Sc., and Ph.D. degrees in signal and information processing from Xidian University, Xi an, China. He is currently
a Full Professor with the School of Electronic Engineering, Xidian University. He has authored or co-authored over 50 scientific articles at top venues, including the IEEE TRANSACTIONS ON IMAGE PROCESSING, the IEEE TRANSACTIONS ON NEURAL NETWORKS AND LEARNING SYSTEMS, the IEEE TRANSACTIONS ON MULTIMEDIA, the IEEE TRANSACTIONS ON SYSTEMS, MAN, AND CYBERNETICS, the IEEE TRANSACTIONS ON CYBERNETICS, the International Conference on Computer Vision, the Conference on Computer Vision and Pattern Recognition, the International Joint Conference on Artificial Intelligence, and the Association for the Advancement of Artificial Intelligence. His research interests include computer vision, multimedia processing and analysis, and information hiding
\end{biography}

\begin{biography}[{\includegraphics[width=1in,height=1.25in,clip,keepaspectratio]{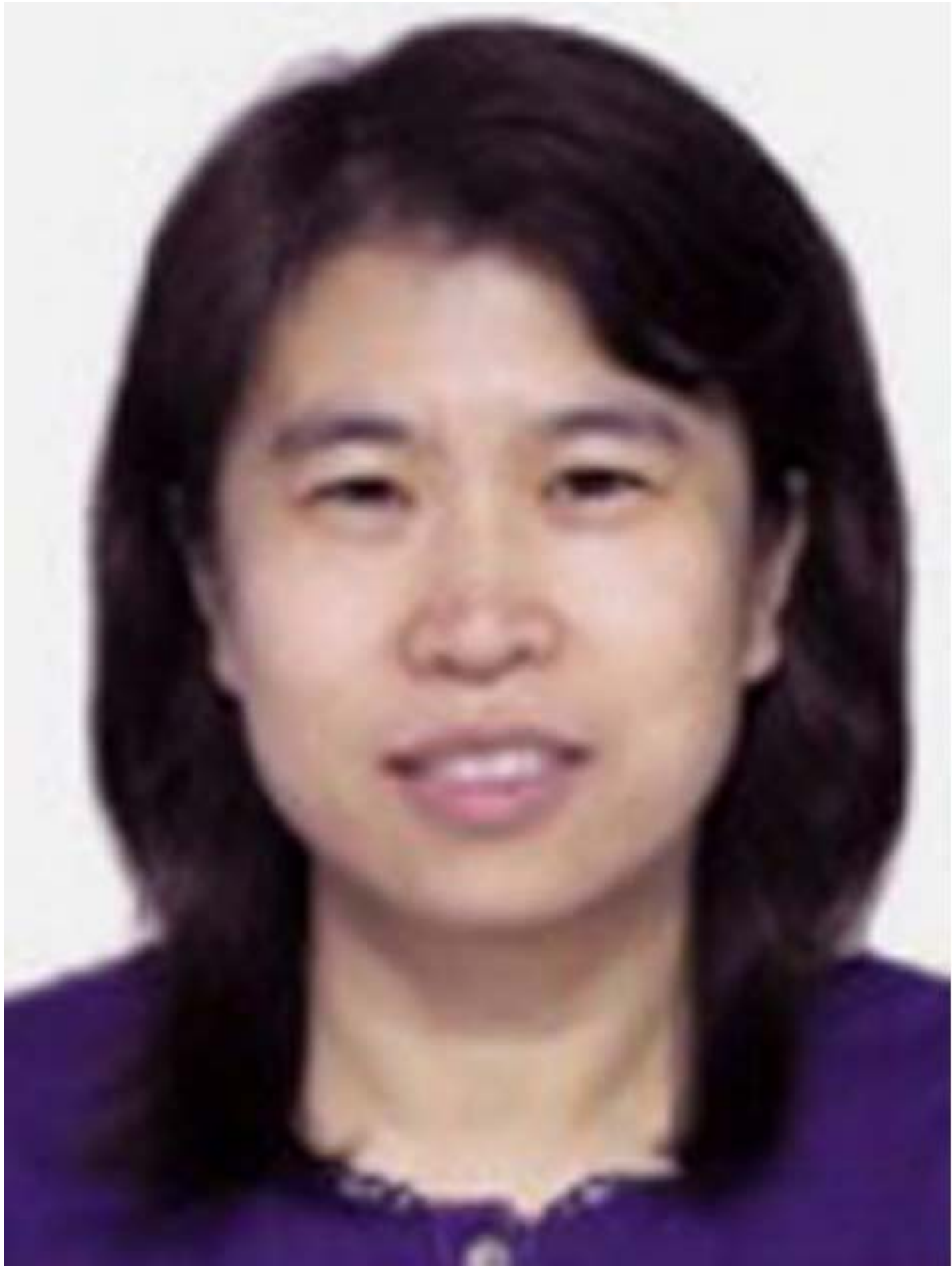}}]{Bo Lang}
received the Ph.D. degree from Beihang University, Beijing, China, 2004. She has been a visiting scholar at Argonne National Lab/University of Chicago for one year. She is currently a Professor in School of Computer Science and Engineering. Her current research interests include data management, information security and distributed computing. As a principle investigator or core member, she has engaged in many national research projects funded by the National Natural Science Foundation, National High Technology Research and Development (863) Program, etc.
\end{biography}

\begin{biography}[{\includegraphics[width=1in,height=1.25in,clip,keepaspectratio]{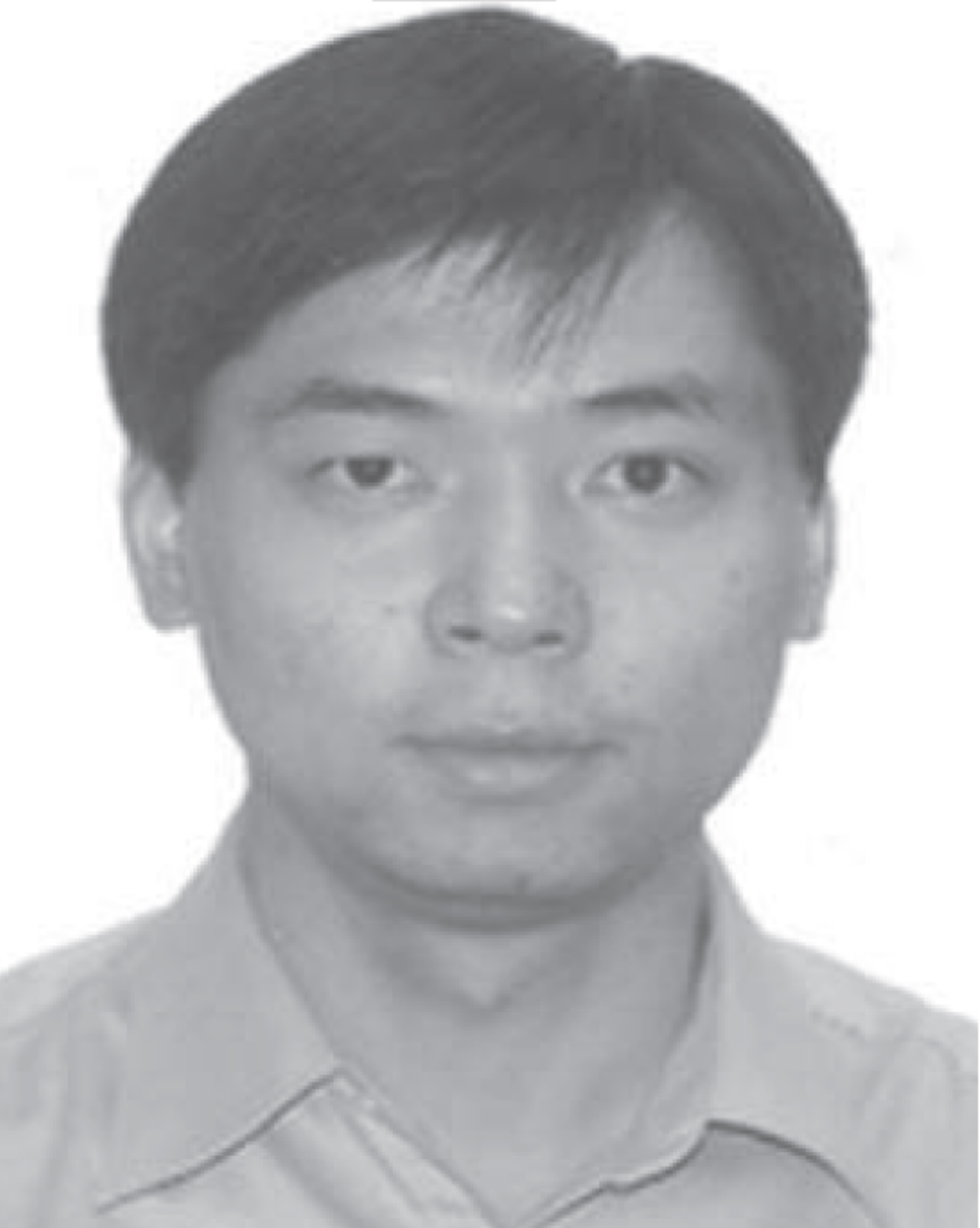}}]{Dacheng Tao} (F'15) is currently a Professor of Computer Science with the Centre for Quantum Computation \& Intelligent Systems, and the Faculty of Engineering and Information Technology, University of Technology, Sydney. He mainly applies statistics and mathematics to data analytics problems. His research interests spread across computer vision, data science, image processing, machine learning, and video surveillance. His research results have expounded in one monograph and over 100 publications at prestigious journals and prominent conferences, such as the IEEE TRANSACTIONS ON PATTERN ANALYSIS AND MACHINE INTELLIGENCE, the IEEE TRANSACTIONS ON NEURAL NETWORKS AND LEARNING SYSTEMS, the IEEE TRANSACTIONS ON IMAGE PROCESSING, the Journal of Machine Learning Research, the International Journal of Computer Vision, the Conference on Neural Information Processing Systems, the International Conference on Machine Learning, the Conference on Computer Vision and Pattern Recognition, the International Conference on Computer Vision, the European Conference on Computer Vision, the International Conference on Artificial Intelligence and Statistics, the International Conference on Data Mining (ICDM), and the ACM Special Interest Group on Knowledge Discovery and Data Mining. He received several best paper awards, such as the Best Theory/Algorithm Paper Runner Up Award in the IEEE ICDM'07, the Best Student Paper Award in the IEEE ICDM'13, and the 2014 ICDM 10 Year Highest-Impact Paper Award.
\end{biography}
\vfill

\end{document}